# Nonlinear regulation of commitment to apoptosis by simultaneous inhibition of Bcl-2 and XIAP in leukemia and lymphoma cells

Joanna Skommer[1], Somkanya C Das[2,§], Arjun Nair[2,§], Thomas Brittain[1], Subhadip Raychaudhuri[2]

[1] School of Biological Sciences, University of Auckland, 3a Symonds Street, Auckland 1142, New Zealand

[2] Department of Biomedical Engineering, University of California, Davis 95616, USA

§ These authors contributed equally

Correspondence:

J. Skommer, E-mail: J.Skommer@auckland.ac.nz

S. Raychaudhuri, E-mail: raychaudhuri@ucdavis.edu

**Abstract**

Background: Apoptosis is a complex pathway regulated by the concerted action of multiple pro- and anti-apoptotic molecules. The intrinsic (mitochondrial) pathway of apoptosis is governed up-stream of mitochondria, by the family of Bcl-2 proteins, and down-stream of mitochondria, by low-probability events, such as apoptosome formation, and by feedback circuits involving caspases and inhibitor of apoptosis proteins (IAPs), such as XIAP. All these regulatory mechanisms ensure that cells only commit to death once a threshold of damage has been reached and the anti-apoptotic reserve of the cell is overcome. As cancer cells are invariably exposed to strong intracellular and extracellular stress stimuli, they are particularly reliant on the expression of anti-apoptotic proteins. Hence, many cancer cells undergo apoptosis when exposed to agents that inhibit anti-apoptotic Bcl-2 molecules, such as BH3 mimetics, while normal cells remain relatively insensitive to single agent treatments with the same class of molecules. Targeting different proteins within the apoptotic network with combinatorial treatment approaches often achieves even greater specificity.

Findings: This led us to investigate the sensitivity of leukemia and lymphoma cells to a pro-apoptotic action of a BH3 mimetic combined with a small molecule inhibitor of XIAP. Using computational probabilistic model of apoptotic pathway, verified by experimental results from human leukemia and lymphoma cell lines, we show that inhibition of XIAP has a non-linear effect on sensitization towards apoptosis induced by the BH3 mimetic HA14-1.

Conclusions: This study justifies further *ex vivo* and animal studies on the potential of the treatment of leukemia and lymphoma with a combination of BH3 mimetics and XIAP inhibitors.

**Introduction**

Oncogenic events, such as genomic instability or oncogene activation, can activate BH3-only proteins and induce the mitochondrial pathway of apoptosis [1]. To counteract these death signals, cancer cells often increase the levels of anti-apoptotic Bcl-2 proteins, and become dependent on them [1]. In such cells agents that mimic the Bcl-2 homology 3 (BH3) domains of the pro-apoptotic Bcl-2 family proteins (BH3 mimetics) induce apoptosis in a single-agent treatment scenario [1]. BH3 mimetics bind in a competitive manner to surface hydrophobic grooves of anti-apoptotic Bcl-2 members thereby releasing the pro-apoptotic Bax/Bak molecules otherwise sequestered in complexes with the anti-apoptotic members [1-3]. Once the threshold of activation of Bax/Bak proteins is reached, the mitochondrial outer membrane becomes permeabilized, leading to the release of cytochrome *c* and other pro-apoptotic proteins into the cytosol. This initiates the assembly of the apoptosome, activation of caspase 9, and then activation of the executioner caspases, such as caspase 3 or 7, which not only directly degrade the proteome and commit cells to death, but also provide feedback cleavage of caspase 9. Inhibitor of apoptosis proteins (IAPs), of which XIAP has been the most widely studied, provide an additional level of regulation, as they inhibit consecutive intermediates (caspase 9 and 3/7) in the apoptotic cascade. Upon release from mitochondria, proteins such as Smac/DIABLO bind to XIAP and free the caspases from inhibition, making another contribution to the activation of effector caspases [4].

Targeted small molecules have a promising future in cancer treatment, as they are potent and highly selective for malignant cells. Nevertheless, high doses of BH3 mimetics have been shown to induce cell death also in normal lymphocytes, narrowing their therapeutic window [5, 6]. The combinatorial treatment with synergistic drugs often achieves greater specificity [7], which led us to investigate the sensitivity of leukemia and lymphoma cells to the pro-apoptotic action of a BH3 mimetic combined with a small molecule inhibitor of XIAP [8].

**Materials and methods**

**Computational modeling**

We developed and studied a Monte Carlo model of the mitochondrial pathway of apoptosis based on our earlier modeling of the same pathway (described in detail in [9,10]). In this stochastic approach we sample diffusion and reaction events of signaling molecules, pertinent to the mitochondrial pathway, using pre-asssigned probability constants. Such probabilistic rate constants are directly estimated from experimentally measured diffusion and kinetic rate constants. Proteins with similar functions are simulated in the model by a single representative protein molecule; for example, Bcl-2 represents all functionally similar anti-apototic Bcl-2 like inhibitors. BH3 mimetic such as HA14-1 and XIAP inhibitor embelin are integrated in the current simulation scheme in such a manner that large number (or concentration) of molecules can be modeled. The advantage of our stochastic simulation approach is that we can study the impact of both cellular variations in protein levels and inherent stochastic fluctuations in signaling reactions [9,10].

**Cell culture and treatments**

Jurkat T, U937, THP-1$\alpha$, CEM and Raji cell lines (ACCT) were maintained and treated in RPMI media (Invitrogen) supplemented with L-glutamine (Invitrogen), 1% penicillin/streptomycin mix (Invitrogen), HEPES (Invitrogen), and 10% fetal bovine serum (FBS; Gibco) at 37 °C in humidified 95% air, 5% $CO_2$. HEK293 cells were cultured in DMEM (Invitrogen), supplemented as above but without the HEPES. Peripheral blood mononuclear cells (PBMCs) were obtained from fresh blood samples of healthy volunteers using standard Ficoll separation. Following isolation, PBMCs were cultured in RPMI media (Invitrogen), supplemented as for other suspension cell lines. Small molecule BH3 mimetic HA14-1 and a small-molecule inhibitor of XIAP (embelin) were procured from Alexis Biochemicals. Both inhibitors were stored in small aliquots and protected from exposure to light. Cells were treated with either of these inhibitors alone, or in combination, using DMSO as a vehicle control.

**Annexin V/7-AAD assay**

To assess cell viability, indicated cell lines were plated on 24-well plates at the density of $0.3 \times 10^6$ cells/ml, and treated as indicated. At the end of the experiment, cells were collected, washed with PBS, and stained for 20 min at RT with Annexin V-PE (Invitrogen) and 7-AAD (Invitrogen; 1μg/sample) in the Annexin V binding buffer (Invitrogen). The cells were analyzed immediately on FACS Calibur (BD). Cells were gated based on forward scatter (FSC) and side scatter (SSC) to exclude cell debris, and next analyzed based on Annexin V-PE fluorescence and 7-AAD fluorescence using CellQuest (BD). Plots were generated using WinMDA.

**Detection of caspase activity**

Activation of caspase 9 was examined by a dual labelling of cells with PhiPhiLux-$G_1D_1$ reagent (OncoImmunin Ltd.), which allows real-time detection of caspase 3 activation, and 7-AAD (Invitrogen), a probe of plasma membrane permeability. Briefly, $1 \times 10^6$ cells were cultured for the time indicated with or without the indicated doses of HA14-1 and/or embelin, harvested, and incubated with PhiPhiLux-$G_1D_1$ following the manufacturer's instructions. Next, cells were stained with 7-AAD for 3 min, and immediately analyzed on a FACS Calibur (BD). Cells were gated based on forward scatter (FSC) and side scatter (SSC) to exclude cell debris, and based on FSC versus 7-AAD to exclude cells with plasma membrane permeability. Plots were generated using WinMDA.

**Results and Discussion**

Using systems level information of the apoptotic signaling reactions (Fig. 1), we have developed a computational model of the mitochondrial pathway of apoptosis, which can be applied to study the activation of caspases over a wide range of apoptotic network parameters (described in detail in [9]). The model explicitly simulates diffusion and reaction events at the level of individual molecules, and is based on a probabilistic method in which the reactivity of all the signaling molecules follows a stochastic, rather than deterministic behavior, probing the induction of apoptosis at a single cell level [10]. At the beginning of the simulation all signaling molecules are distributed randomly and uniformly in the cell volume simulated by a three dimensional cubic lattice. Cytochrome *c* molecules are confined within a fixed mitochondrial region inside the cell volume. Once the concentration of active Bax dimers reaches a pre-assigned threshold value (~ 0.017 μm), cytochrome *c* is released from the mitochondria into the cytosol in an all-or-none manner [11]. The proteins with similar biochemical activities are represented by single species in our simulations. For example, Bax represents all pro-apoptotic multi-domain Bcl-2 proteins (i.e. both Bax and Bak), whereas Bcl-2 represents all anti-apoptotic Bcl-2 proteins (e.g. Bcl-2, Bcl-$X_l$, Mcl-1, etc.). The model uses truncated Bid (tBid), enhanced by the action of BH3 mimetic HA14-1, as a trigger for the Bax activation in the mitochondrial pathway of apoptosis. With this approach we have previously shown that Bcl-2 regulates cell-to-cell variability in time-to-death, and that stochastic fluctuations in apoptotic reactions are sufficient for survival of single cells upon treatment with a BH3 mimetic [9]. In single cells, mitochondrial stress is translated into a binary decision at the level of caspase activation, which is subject to further regulation by feedback loops, e.g. involving XIAP, caspase 9 and caspase 3 (Fig. 1). Such a feedback loop can be used to suppress weak post-mitochondrial stress signals, increasing cell-to-cell variability [12-14]. Increased XIAP expression has been reported in a variety of human tumors as well as in leukemic cell lines, and XIAP over-expression (or higher XIAP/Smac ratio) has been linked to poor prognosis [15,16]. In our computational model a two-fold increase in the average value of XIAP, from 18 (corresponding to 30nM concentration, as found in normal cells) to 36 (corresponding to 60nM concentration), did not have a significant effect on inhibition of apoptosis (Fig. 2a). A ten-fold increase in the average XIAP level, however, had a strong inhibitory effect on cell death activation (Fig. 2a). Ten-fold increase in XIAP level could also effectively reduce the threshold Bcl-2 level needed to suppress apoptotic death, a strategy that can be used by cancer cells to overcome their inherent vulnerability to apoptosis caused by oncogenic activation of pro-apoptotic BH3-only proteins. Our computational model of apoptosis (tBid mediated) indicated that the effect of XIAP on inhibition of sensitivity to cell death has a non-linear nature due to coupling of XIAP to several signalling molecules downstream of mitochondria, which generates a loop structures (Fig. 1, grey box). When Bcl-2 and XIAP are inhibited by HA14-1 and embelin, respectively, in a combined treatment, effective induction of cell death is observed at a lower dose of HA14-1 (Fig. 2b). Further increase in embelin concentration did not significantly increase apoptotic death induced by varying doses of HA14-1, again underscoring the non-linear pro-apoptotic effect of the combinatory treatment (Fig. 2b).

To validate the predictions of the mathematical modeling we analyzed the induction of apoptosis in Jurkat T cells treated with increasing concentrations of the BH3 mimetic HA14-1 (Axxora), alone or in combination with embelin (Sigma). Apoptotic cell death was measured using Annexin V and 7-AAD (both from Invitrogen) and flow cytometry (FACSCalibur, BD) (Fig. 3a). Administration of embelin significantly enhanced apoptotic cell death induced by HA14-1 (Fig. 3b). Importantly, while cell death induced by low doses (5-10μM) of HA14-1 increased significantly upon co-treatment with embelin, this effect was not observed at higher doses of HA14-1 (Fig. 3b), confirming the nonlinear effect of XIAP inhibition as suggested by our computational modeling. We also studied cell lines derived from other hematological malignancies, including monocytic leukemia (THP-1α), histiocytic lymphoma (U937), T-cell acute lymphoblastic leukemia (CEM), and B-cell lymphoma (RAJI), and observed again that embelin enhances significantly cell death induced by low doses (5-10μM) of HA14-1 (Fig. S1 and data not shown). Using a fixed-ratio analysis of the pharmacological interactions between HA14-1 and embelin we observed that co-administration of the two compounds is synergistic [17] at the doses of HA14-1 that induce only limited cell death in a single-agent treatment, independently of the cell line used (Fig. 3c). However, administration of embelin did not sensitize to cell death induced by higher doses of HA14-1 (Fig. 3c), again confirming the non-linear effect of XIAP inhibition on sensitivity to the BH3 mimetic. Importantly, non-cancerous HEK293 cells, which are resistant to cell death induced by HA14-1 [9], were not sensitized by embelin (3d). Similarly, in peripheral blood mononuclear cells (PBMCs), obtained from healthy individuals, treatment with embelin did not enhance HA14-1-induced apoptosis (Fig. 3e).

We showed previously that the combined effects of (a) all-or-none activation of caspase 9 and 3 at the single cell level, and (b) cell-to-cell fluctuations lead to bi-modal probability distributions for activated caspase 9 and 3 [9]. This study shows that the combined effect of higher Bcl-2 and XIAP levels leads to efficient suppression of cell death, and results in an increase in the time-to-death and its cell-to-cell variability. In a combined HA14-1-embelin treatment scenario, increase in embelin dose resulted in decreased cell-to-cell variability in time-to-death but the characteristic all-or-none type activation remained preserved (Fig. 4). Such all-or-none type activation with cell-to-cell variability leads to bi-modal probability distributions in caspase-3 activation [9,10], which can be used for quantitative estimation of cell-to-cell variability in apoptosis activation [Fig. 5a]. Direct estimation of cell-to-cell variability in time-to-death, such as its variance, is problematic because only a fraction of cells undergoes apoptosis at a given time. Assuming perfect bi-modal curves, the ratio variance/average for caspase 3 activity is found to be C3[1-f(e,t)], where C3 is the initial number of Caspase 3 molecules and f(e,t) denotes the fraction of cells in which caspase 3 has undergone complete activation (for a given embelin concentration e at time t) [9]. Estimation of C3[1-f(e,t)] from Fig. 4 (at the end of simulation, $t = 5 \times 10^4$ seconds): 0.81C3 (embelin = 200 molecules, ~ 0.33 μM ), 0.44C3 (embelin = 1000 molecules, ~1.67 μM ) and 0.37C3 (embelin = 2000 molecules, ~3.33 μM ), which indicates a decrease in cell-to-cell variability due to increased XIAP inhibition (by increased concentration of embelin). In our simulations, a cellular average of C3 = 60 molecules (~ 100 nm) is used for inactive caspase 3 molecules. Estimation of variance/average from the experimental data (early apoptotic cells in Fig. 3b) also shows decreased cell-to-cell variability with increased XIAP inhibition: 0.91C3 (embelin = 0), 0.83C3 (embelin = 10 μM), 0.66C3 (embelin = 20 μM) for HA14-1 = 5 μM.

To validate the findings of the computational model, we analyzed the distribution of caspase 3 activity in early apoptotic cells using caspase 3 recognition sequence DEVDGI labeled with a fluorophore (PhiPhiLux-$G_1D_1$) and 7-AAD (a marker of early plasma membrane permeabilisation). Treatment with HA14-1 led to increased caspase 9 activity, with a clearly distinguishable population of cells with over 10-fold increase in PhiPhiLux-$G_1D_1$ fluorescence (Fig. 5b). The intermediate events in terms of caspase 3 activity were infrequent (<8%; M2 on Figure 5b), in line with our previous observations [8]. Upon addition of inhibitor of XIAP embelin there was an approximately two-fold increase (up to 15%) in the number of intermediate caspase 3 activation events (Fig. 5b). This confirms that direct inhibition of XIAP, and thus interference in feedback loops that regulate activation of caspases downstream of mitochondria, can somehow disturb the bimodal distribution of caspase 3 activity. Considering the possibility of non-specific labeling with reagents such as PhiPhiLux, which contain an artificial caspase cleavage site, as well as varying sensitivity of different approaches for detection of caspase activity, it will be of interest to perform similar analysis using alternative experimental avenues that allow assessment of caspase activity in living cells, for example with the use of recently described crown nanoparticle probes [18].

The advantage of *in silico* models of cellular systems is that one can selectively study the behavior of specific pathways, reducing the complexity of biological systems, and thus test concepts that are otherwise difficult to verify experimentally [10, 19]. Recent developments in the field of computational systems biology have paved the way to important new discoveries pertinent to cancer therapy [9, 20, 21]. Mathematical approaches range from models of cancer epidemiology [22] or models reproducing the dynamics of accumulation of genetic changes during tumorigenesis [23], through computational tools for predicting response of cancer cells to treatment [24], up to early stage models for design of patient-specific immunotherapy [25], and all provide invaluable insights into cancer biology. Considering the importance of apoptotic cell death in tumorigenesis as well as cancer therapy, a detailed systems-level understanding of this pathway is also likely to yield clinically-relevant information. Even though there are some discrepancies between the behavior of the computational model and the real cellular systems, which may be due to the off-target effects of the BH3 mimetic *in vitro*, particularly at higher concentrations, approximate quantitative data for some protein species in the computational model, or the fact that the combination of tBid and HA14-1 is used in our model, the general behaviour of our model and biological system follows the same trend.

Our network analysis suggests that once the anti-apoptotic reserve of a cell is overwhelmed by inhibition of both Bcl-2 and XIAP, a higher sensitivity of cancer cells to small doses of BH3 mimetics is observed. Our computational model of apoptosis suggested a nonlinear effect of XIAP inhibition, which we verified experimentally in several leukemia and lymphoma cell lines. Such non-linear effect of XIAP, expression of which is regulated by cytokines and other survival factors through the PI3K/MAPK pathway [16], can provide a mechanism for unusual apoptosis resistance of cancer cells. Therefore, a combination of low doses of BH3 mimetics with a pharmacological inhibitor of XIAP embelin represents a particularly promising strategy for the treatment of hematological malignancies. Our computational studies indicate that combined inhibition of both Bcl-2 and XIAP can reduce cell-to-cell variability in cell death and thus can reduce fractional killing of tumor cells [9]. This study seems to hold sufficient promise to justify further *ex vivo* and animal studies on the potential of the treatment of leukemia and lymphoma with a combination of BH3 mimetics and embelin.

**Authors' contribution**

JS and TB carried out the experimental studies. SR, SD, and AN designed and carried out the computational studies. JS and SR conceived the study and drafted the manuscript. TB, SD, and AN helped in preparation of the manuscript. All authors read and approved the final manuscript.

**Acknowledgments**

We are grateful to Prof. R. Dunbar (SBS, University of Auckland) for providing U937 and peripheral blood mononuclear cells, Prof B. Baguley (UoA) for providing Raji CEM and HL-60 cells, Dr J. Taylor (SBS, UoA) for THP-1α cells, and Dr D. Wlodkowic (Department of Chemistry, UoA) for providing PhiPhiLux-$G_1D_1$ reagent.

**Figure 1** Schematic of the mitochondrial pathway network used in the probabilistic computational model. MOMP, mitochondrial outer membrane permeabilisation; XIAP, X-linked inhibitor of apoptosis protein. The model emulates the experimental set up by using HA14-1 as a stress trigger, and embelin as XIAP inhibitor. Binding affinity of HA14-1 to Bcl-2 and embelin to XIAP were based on previously published data (8, 26). $K_D$, dissociation constant.

**Figure 2** Probabilistic computational model of apoptosis suggests that inhibition of XIAP leads to a nonlinear increase in sensitivity to apoptosis. Data was obtained after $5 \times 10^8$ Monte Carlo (MC) simulation steps. 1MC step = $10^{-4}$ seconds.
(A) The percentage of apoptotic cells is shown for various levels of Bcl-2 and XIAP. We did not observe any activation for Bcl-2 = 2250 (number of molecules) and XIAP = 180 (number of molecules). XIAP (or Bcl-2) concentration (in nanomolar) was 1.67 times the number of molecules used in our simulations.
(B) Increasing apoptotic stress (combined dose of HA14-1 and embelin), with Bcl-2 = 2250 (number of molecules) and XIAP = 180 (number of molecules). Concentration for a molecular species can be obtained by multiplying the number of molecules with a factor of 1.67.

**Figure 3** Embelin is synergistic with low doses of HA14-1 in leukemia cells, but not in normal cells
(A) Jurkat T cells were treated with increasing concentration of HA14-1, alone or with embelin, and the percentage of viable, early and late apoptotic cells was determined by flow cytometry. Cells negative for Annexin V/7-AAD staining (R2) are considered as viable; Annexin V positive cells are considered as early apoptotic (R3) or late apoptotic (R4), depending on the permeability to 7-AAD.
(B) Data from 3 independent experiments on Jurkat T cells are shown as mean ± SEM. [a] $p<0.05$ compared to HA14-1 5μM; [b] $p<0.05$ compared to HA14-1 10μM
(C) Indicated cells were cultured in the presence of escalating doses of HA14-1 (0-25μM) or embelin (0-70μM), used for the calculation of the theoretical dose-response surface in combinatory treatments, or combinations of the 2 agents at a 1:2 molar ratio (5/10, 7.5/15, 10/20, 12.5/25, 15/30, 20/40). The percentage of apoptotic cells was determined after 24h using flow cytometry. The results were analyzed using the CombiTool software (9, 18). The experimental data ( ● ) was compared with the theoretical dose–response surface (○) representing the calculated additive effect of the combined doses of HA14-1 and embelin. If the experimental data points are mapped above the theoretical surface, coincided with it, or were below it, the interaction of modalities is defined as synergistic, additive, or antagonistic, respectively. The computed experimental data is the mean of 3 independent experiments.

(D) HEK293 cells were treated with DMSO or HA14-1 (25μM), with or without embelin (30μM), and after 24h their viability was analyzed as in a). Data from 3 independent experiments are shown as mean ± SEM.

(E) Percentage of apoptotic cells after the indicated treatment of peripheral blood mononuclear cells from healthy individuals, assessed as in (a). Data from a representative sample out of 3 analyzed is shown.

**Figure 4** Effects of XIAP inhibition on the cell-to-cell variability in caspase activation.

Time course of capsase-3 activation (normalized to 1) is shown as obtained from the computational model. Time is shown in Monte Carlo (MC) simulation steps up to $5 \times 10^8$ MC steps (1 MC step = $10^{-4}$ seconds). Three Embelin values (top to bottom) are used in our simulations: 200, 1000, and 2000 (number of molecules) and HA14-1 = 1800 molecules (concentration ~ 3 μM). Number of molecules of Bcl-2 = 2250 and XIAP = 180. Data is shown for 15 individual single cell runs (only a fraction of cells show activation within the given simulation time).

**Figure 5** Effects of XIAP inhibition on the bi-stable mode of caspase activation in HA14-1-treated cells

(A) Probability distribution of caspase-3 activation at time steps = $10^4$, $2 \times 10^8$ and $5 \times 10^8$ (left to right). Time is shown in Monte Carlo (MC) simulation steps (1 MC step = $10^{-4}$ seconds). Three Embelin values (top to bottom) are used in our simulations: 200, 1000, and 2000 (number of molecules). Number of molecules of Bcl-2 = 2250, XIAP = 180, and HA141-1 = 1800. Concentration for a molecular species can be obtained by multiplying the number of molecules with a factor of 1.67. Data is obtained from 16 individual single cell runs.

(B) Jurkat T cells were treated as indicated (embelin 20μM) for 24h, stained with PhiPhiLux-$G_1D_1$, and analyzed by flow cytometry. M1, full caspase activation; M2, intermediate caspase activation; M3, no caspase activation. The population of 7-AAD negative cells was analyzed to exclude for potential loss of caspase labeling during the later stages of cell death.

**Figure 1**

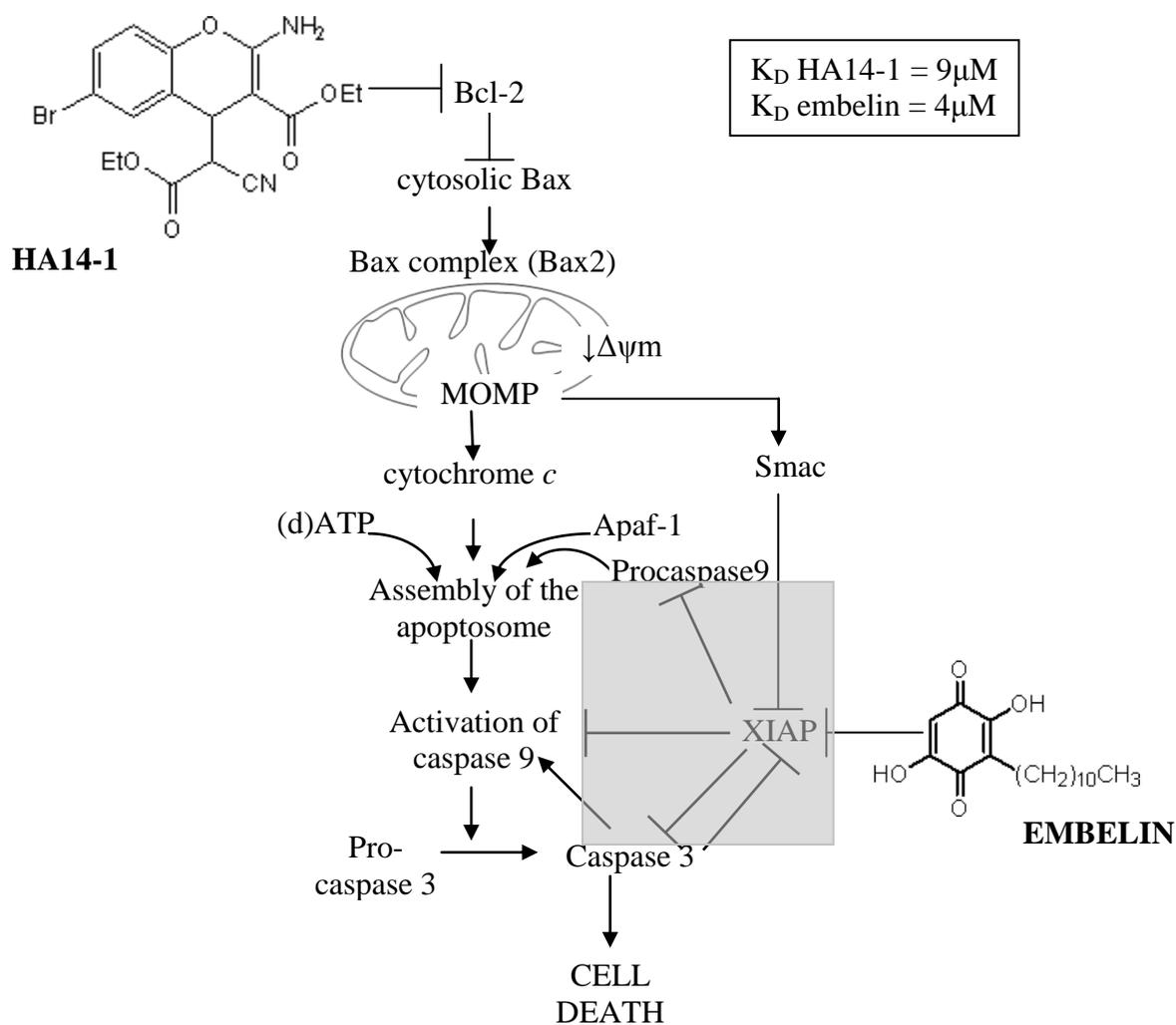

Figure 2

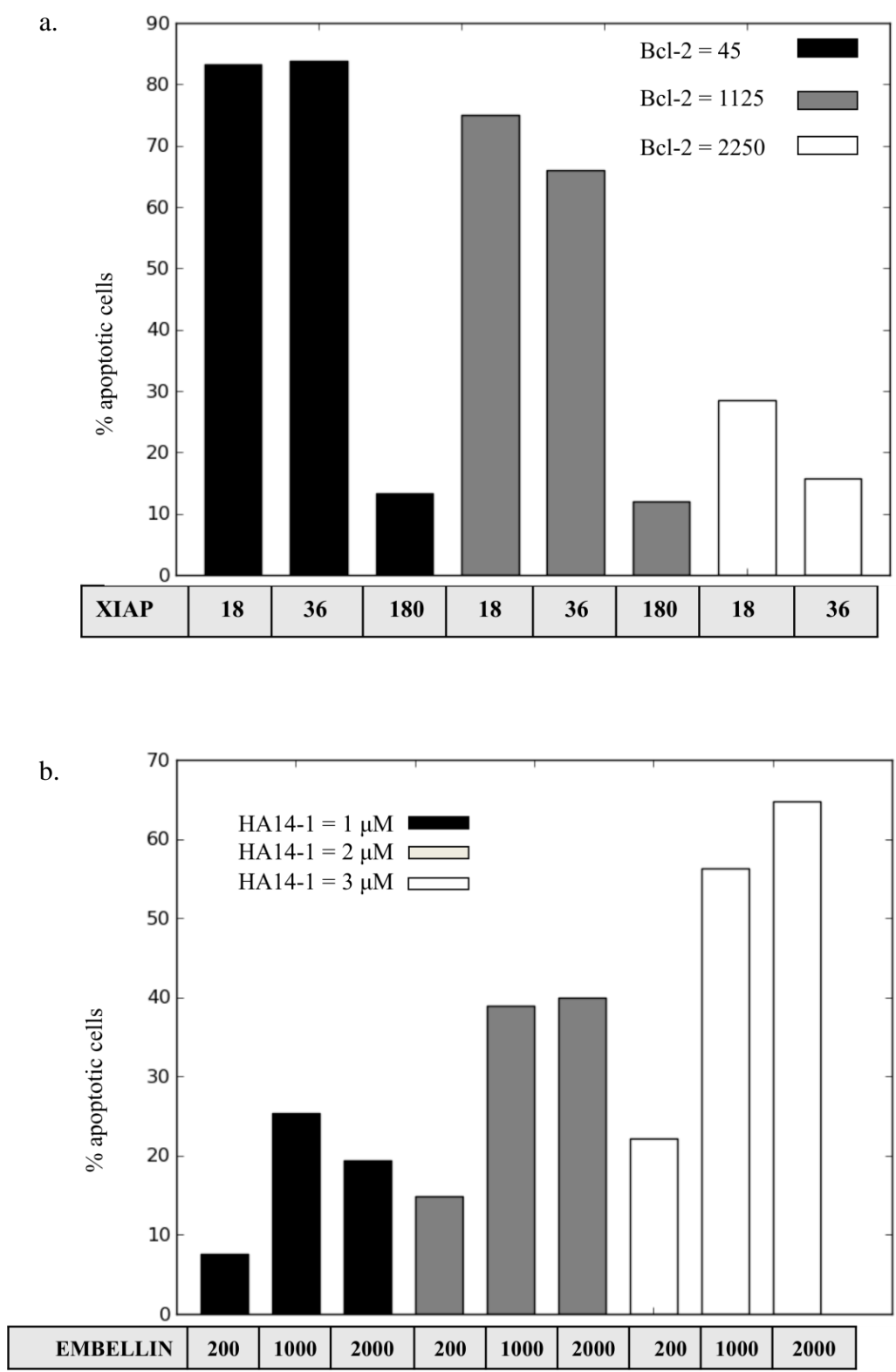

**Figure 3**
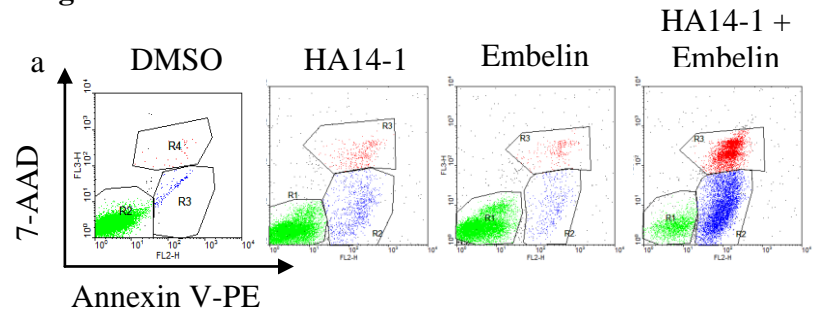
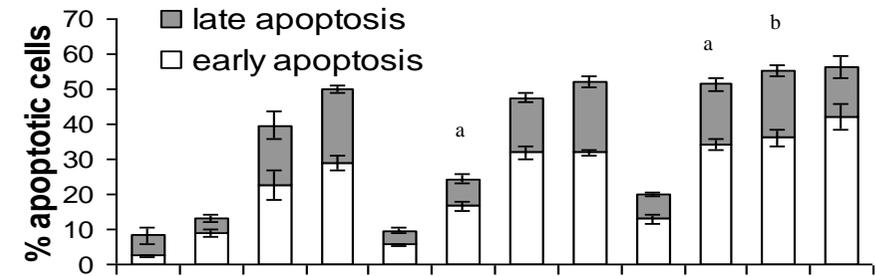
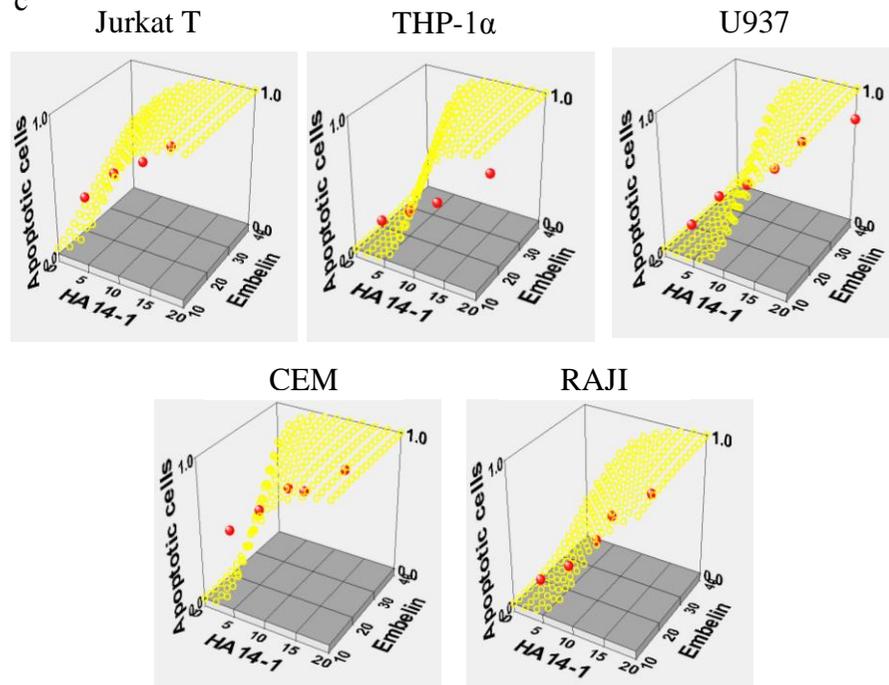
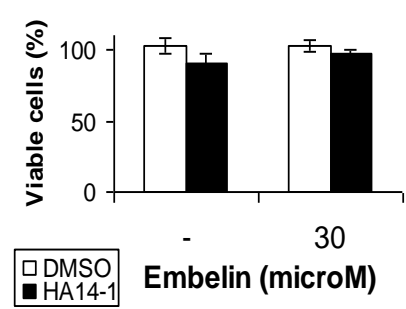
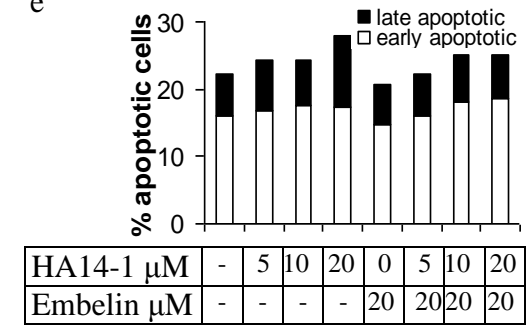

Figure 3

Figure 4

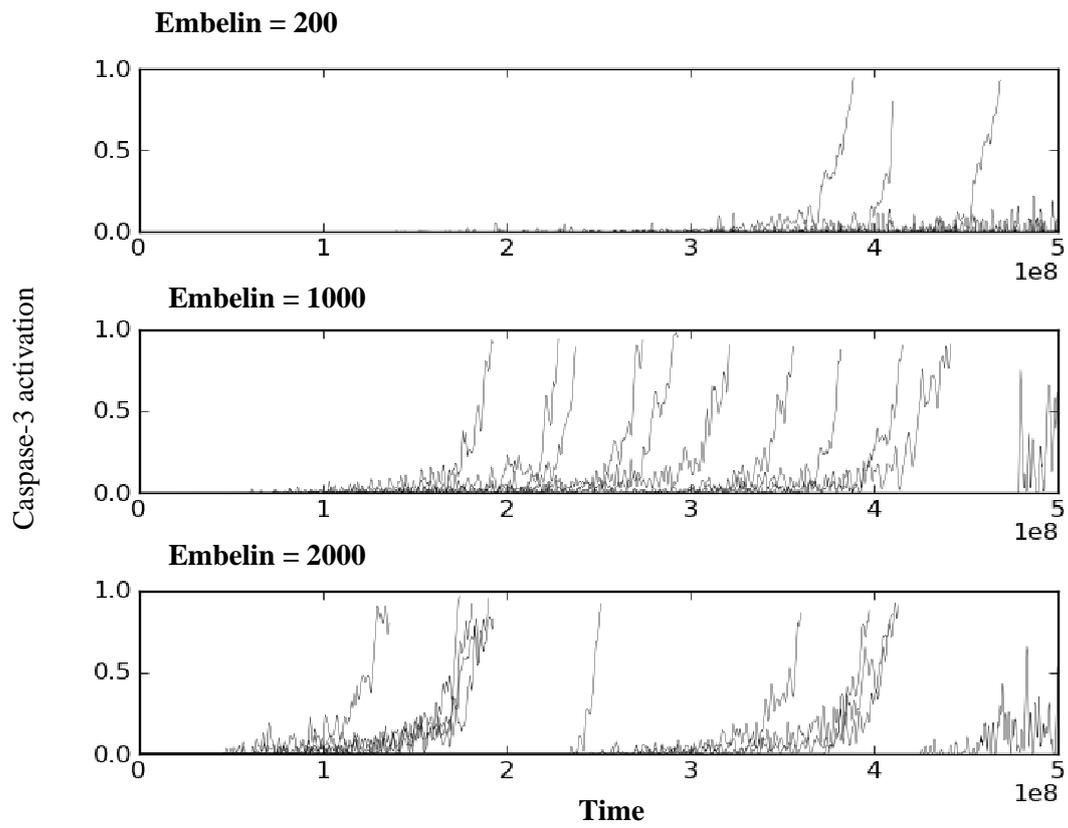

Figure 5

a.
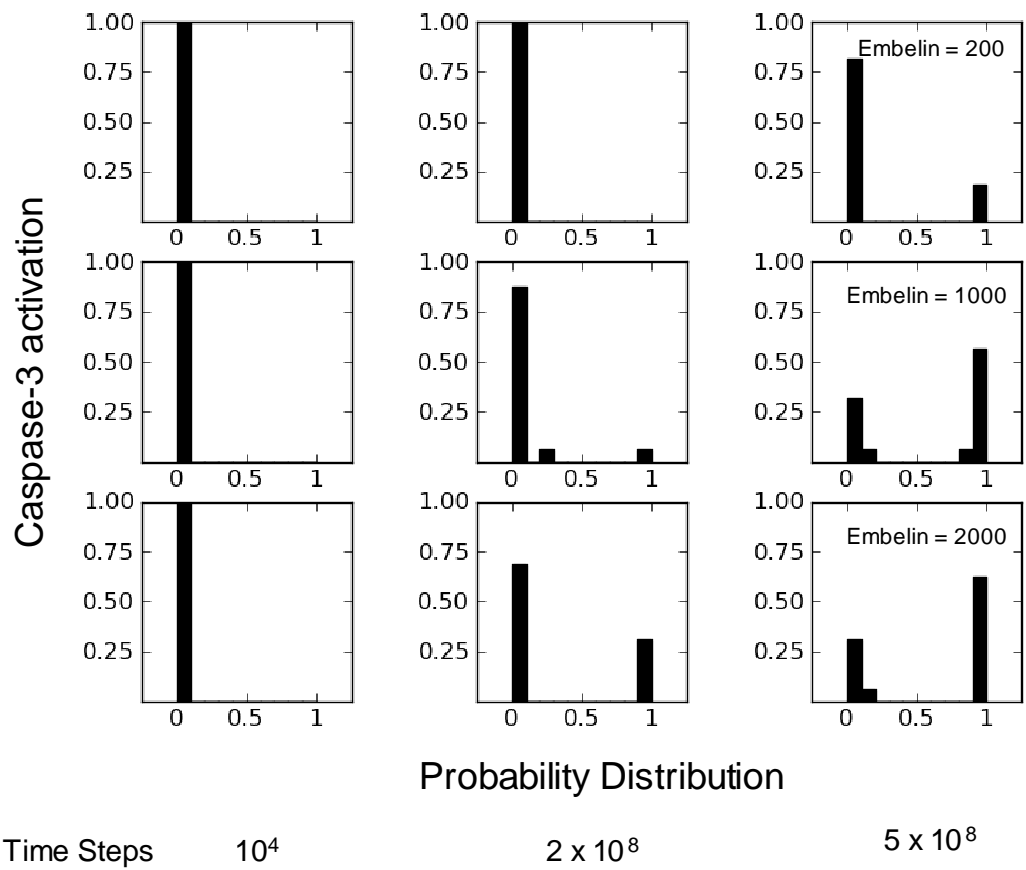

b.
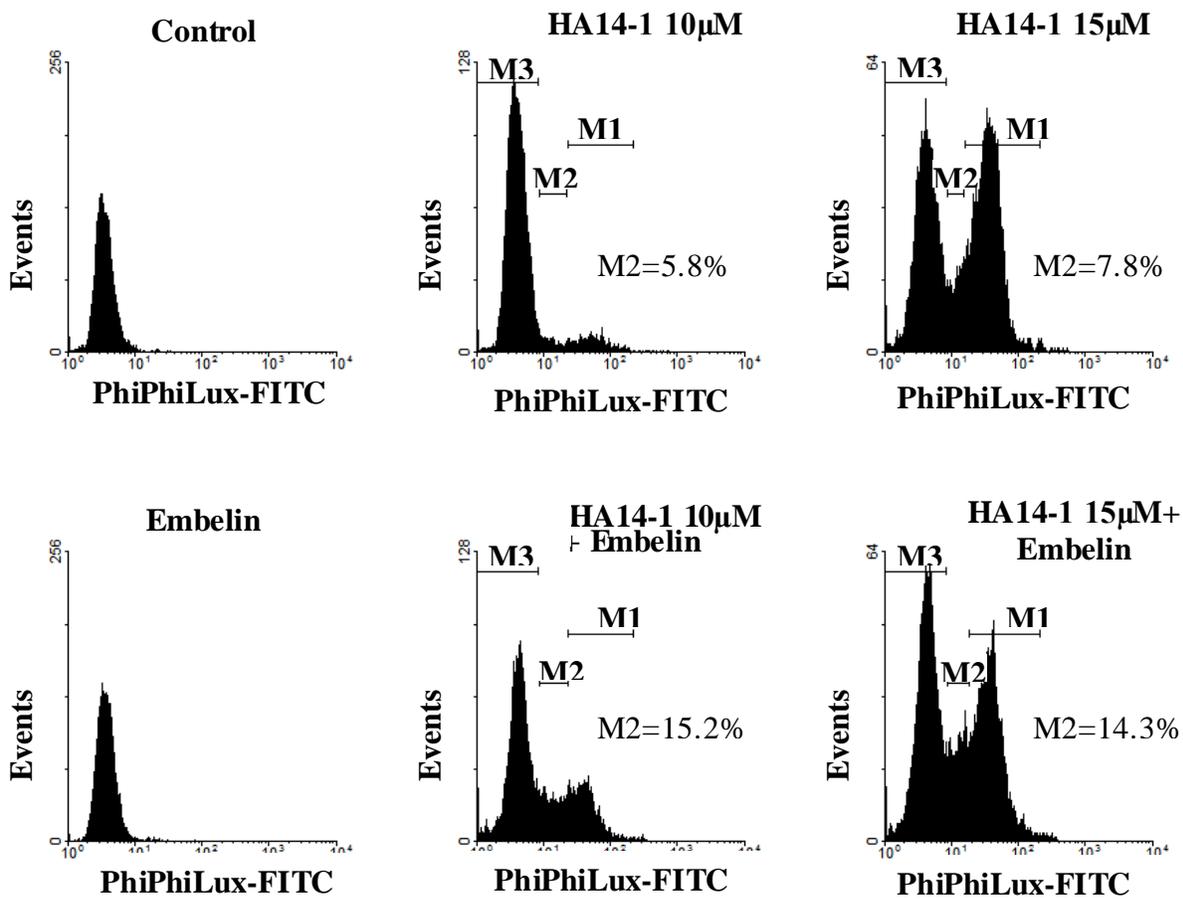

**Supplementary Figure**
Click here to download Figure: Supplementary Figure 1.doc

Supplementary Figure 1

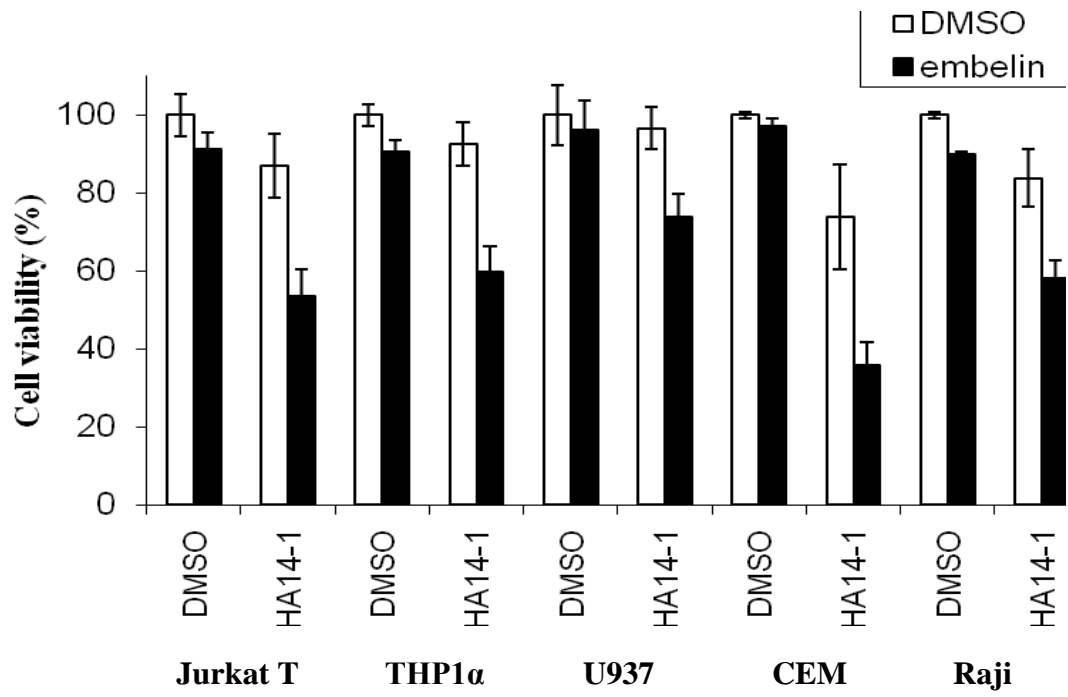